\documentclass[letterpaper]{jpconf}
\begin{document}

\title{The Sanford Underground Research Facility}


\author{J Heise}

\address{Sanford Underground Research Facility, 630 East Summit Street, 
Lead, SD 57754}

\ead{jaret@sanfordlab.org}

\begin{abstract} 
The former Homestake gold mine in Lead, South Dakota, has been transformed 
into a dedicated facility to pursue underground research in rare-process 
physics, as well as offering unique research opportunities in other 
disciplines. The Sanford Underground Research Facility (SURF) includes two 
main campuses at the 4850-foot level (4300 m.w.e.) -- the Davis Campus and 
the Ross Campus -- that host a range of significant physics projects: the 
LUX dark matter experiment, the {\sc Majorana Demonstrator} neutrinoless 
double-beta decay experiment and the CASPAR nuclear astrophysics 
accelerator. Furthermore, the BHUC Ross Campus laboratory dedicated to 
critical material assays for current and future experiments has been 
operating since Fall 2015. Research efforts in biology, geology and 
engineering have been underway at SURF for 10 years and continue to be a 
strong component of the SURF research program. Plans to accommodate future 
experiments at SURF are well advanced and include geothermal-related 
projects, the next generation direct-search dark matter experiment 
LUX-ZEPLIN (LZ) and the Fermilab-led international Deep Underground 
Neutrino Experiment (DUNE) at the Long Baseline Neutrino Facility (LBNF). 
SURF is a dedicated research facility with significant expansion 
capability, and applications from other experiments are welcome.
\end{abstract}


\section{Introduction}

The Sanford Underground Research Facility (SURF) is a unique facility 
built on rich legacies in both mining and transformational science to 
serve a variety of research disciplines, including physics, biology, 
geology and engineering~\cite{SURF-Heise-LRT2015,SURF-Heise-2015}.  In 
particular, a deep underground laboratory enables investigations into some 
of the most fundamental topics in physics today, including the nature of 
dark matter, the properties of neutrinos and the synthesis of atomic 
elements within stars.

Opened July 2007, SURF just celebrated 10 years as a dedicated science 
facility, which would not have been possible without generous donations 
from the Barrick Gold Corporation (owner of the Homestake Mining Company), 
South Dakota philanthropist, T.\ Denny Sanford, as well as strong support 
from the South Dakota legislature.  Since 2012, SURF operation has been 
funded by the U.S.\ Department of Energy through sub-contracts with various 
national laboratories (initially LBNL and since October 2016 via 
Fermilab).

In total, the SURF facility consists of more than 600~km of tunnels and 
shafts extending from the surface to over 2450~meters (8000~feet) below 
ground.  Of the 29 underground elevations that are currently accessible, 
six have been identified as key levels for science activities: 300L, 800L, 
1700L, 2000L, 4100L, 4850L.  The Laboratory property comprises 223~acres 
(expanded in May 2017) on the surface and 7700~acres underground, and the 
Surface Campus includes approximately 26,088 gross square meters in 
structures (expanded in May 2017) that were for the most part inherited 
from Barrick.



\section{Facility Operations Infrastructure}

Maintenance and operation of key elements of facility infrastructure 
enable safe access underground.  Two main shafts -- the Ross and Yates -- 
provide redundant routes for power and network services as well as for the 
transportation of personnel and materials.  Additional shafts are 
dedicated to ventilation, providing air-flow at a rate of approximately 
510,000 m$^{3}$/hr, with over 40\% directed to the 4850L.  Air-flow is 
expected to increase significantly to support LBNF excavation expected to 
begin mid-2019.  Various air handling units and chiller systems provide 
the required air flow inside laboratory spaces as well as condition the 
environment to meet experiment temperature and relative humidity 
requirements.

To provide increased capacity to support future experiments, extensive 
renovations are being performed in the Ross shaft. Since beginning in 
August 2012, new steel supports and associated ground support have been 
installed through roughly 97\% of the total length.  The entire upgrade 
project is expected to be completed in 2018.

\section{Surface Science Facilities}

SURF facilities support research activities both on the surface as well as 
underground.  On the surface, science activities are facilitated in a 
number of ways, but the principal facility that directly serves science 
needs is the Surface Laboratory, which provides approximately 210~m$^{2}$ 
of lab space (265~m$^{2}$ total) in the top-most level of a four-story 
building. The Surface Laboratory facility includes a cleanroom that was 
installed in 2009 (37~m$^{2}$, originally used by LUX) as well as new 
systems installed in 2017 to support LUX-ZEPLIN (LZ) detector assembly 
activities, including a new metal, low-radon cleanroom (55~m$^{2}$) served 
by a radon-reduction system fabricated by Ateko capable of providing 
air-flow at a rate of approximately 300~m$^{3}$/hr.


The main infrastructure for the support of science activities underground 
has been developed on the 4850-foot level (4300 meters water equivalent) 
with multi-laboratory campuses located near both the Ross and Yates 
shafts. Near the Yates shaft, a laboratory complex called the Davis Campus 
has been operating since 2012. It has a footprint (1015~m$^{2}$ science, 
3017~m$^{2}$ total) consisting of clean laboratory spaces that are 
typically maintained around Class 3000 or lower.  The 4850L Ross Campus 
includes four existing excavations that were used as maintenance shops 
during mining activities with a footprint consisting of 1148~m$^{2}$ 
(science) and 2645~m$^{2}$ (total).  Initial occupancy of the Ross Campus 
was in 2011 followed by new laboratories in 2015.

\section{Laboratory Characterization}

A geologic model has been constructed to incorporate the complex surface 
topology as well as the six main geologic formations plus other features 
that characterize the underground environment.  Previously 
published~\cite{SURF-Heise-2015} overburden density values were calculated 
using several discrete points.  Results from a more sophisticated analysis 
using the relative rock formation volumes weighted by the average 
formation densities for a variety of cone angles are summarized in 
Table~\ref{tab:density}.

\begin{table}[htbp] 
\caption{\label{tab:density}Overburden rock density estimates at different 
cone angles for various underground locations at SURF, using a 
3-dimensional geological model~\cite{RoggenthenHart-2014}.  Angles are 
relative to the vertical above the specific site.  Overburden 
density errors are estimated to be $<$1\%.}
\begin{center} 
\begin{tabular}{lcccccc} \br {\bf Location} & {\bf 
Rock} & \multicolumn{5}{c}{\bf 
Overburden Density} \\
 & {\bf Overburden} & {\bf 0 deg} & {\bf 15 deg} & {\bf 30 deg} & {\bf 45 
deg} & {\bf 60 deg}\\
 & {\bf (m.w.e)} & \multicolumn{5}{c}{\bf (g/cm$^3$)} \\ 
\mr
\multicolumn{6}{l}{\bf 4850L Davis Campus} \\
\mr
LUX/LZ Detector  & 4210 & 2.870 & 2.898  & 2.848  & 2.833  & 2.828 \\ 
MJD Detector     & 4260 & 2.882 & 2.892  & 2.848  & 2.832  & 2.828 \\ 
\mr
\multicolumn{6}{l}{\bf 4850L Ross Campus} \\
\mr
MJD Electroforming  & 4290 & 2.853  &&&& 2.821 \\ 
BHUC                & 4380 & 2.916  &&&& 2.821 \\ 
CASPAR              & 4170 & 2.783  &&&& 2.820 \\ 
\mr
\multicolumn{6}{l}{\bf 4850L LBNF Campus} \\
\mr
Chamber 1    & 3980 & 2.791  &&&& 2.818 \\ 
Chamber 2    & 3860 & 2.755  &&&& 2.817 \\ %
Chamber 3    & 3810 & 2.760  &&&& 2.816 \\ 
Chamber 4    & 3830 & 2.782  &&&& 2.815 \\ 
\mr
\multicolumn{6}{l}{\bf Other} \\
\mr
800L (Muon site~\cite{Bkgd-Muon})  &  770 & 2.729  &&&& 2.717 \\ 
2000L (Muon site~\cite{Bkgd-Muon}) & 1700 & 2.725  &&&& 2.754 \\ 
\br
\end{tabular}
\end{center}
\end{table}

SURF and other groups have collected data characterizing the facility in 
terms of various radioactive backgrounds.  The Davis Campus is hosted in 
Yates Amphibolite rock, which is relatively low in radioactivity: 0.22~ppm 
U, 0.33~ppm Th and 0.96\% K.  The Poorman rock formation surrounding the 
Ross Campus is slightly higher in natural radioactivity: 2.58~ppm U, 
10.48~ppm Th and 2.12\% K~\cite{Assay-Whitepaper, Assay-Oroville}.  
Long-term underground radon data have been collected at various locations.  
In particular, the total average radon concentration over the monitoring 
period at the Davis Campus (1936 days) is approximately 300~Bq/m$^{3}$, 
with a low baseline of 150~Bq/m$^{3}$.  For the same period, the average 
radon concentration at the Ross Campus is approximately 
500--600~Bq/m$^{3}$.  Brief excursions above 1000 Bq/m$^{3}$ have been 
observed at both campuses, typically correlated with maintenance and 
ventilation changes.  Other efforts to characterize physics backgrounds in 
various underground areas were carried out by various research groups: 
muons~\cite{Bkgd-Muon, Bkgd-Muon_MJD}, neutrons~\cite{Bkgd-Neutron_Best} 
and gamma rays~\cite{Bkgd-Gamma}.

\section{Current Science Program}

As interest grows from the scientific community, the formal process for 
implementing experiments at SURF has also matured~\cite{SURF_science}. To 
date, thirty-eight groups have conducted research programs at SURF, 
including efforts at elevations ranging from the surface to the 5000L.  A 
total of twenty-three research groups are currently active.

The Large Underground Xenon (LUX)~\cite{LUX-Complete} experiment performed 
a direct search for dark matter at the 4850L Davis Campus using 370~kg of 
xenon housed within an ultrapure titanium cryostat, which was immersed in 
an ultrapure 270-tonne water shield.  The LUX detector stopped data 
collection in May 2016 and held world-leading sensitivity for $\sim$3.5 
years over most of the WIMP-mass region.  The LUX experiment was 
decommissioned in 2017.

The {\sc Majorana} collaboration is investigating neutrinoless double-beta 
decay at the 4850L Davis Campus using the {\sc Majorana Demonstrator} 
(MJD) detector~\cite{MJD,MJD-TAUP2017}, which consists of 44~kg of 
germanium detectors (approximately 30~kg enriched in $^{76}$Ge) within two 
ultrapure copper cryostats protected by a 66-tonne shield comprised of 
layers of copper, lead and HDPE with an active muon veto. The {\sc 
Majorana} group is currently transitioning from commissioning to 
production operation and plans to continue operations through 2020. The 
{\sc Majorana Demonstrator} electroforming facility that operated at the 
Ross Campus since 2011 was decommissioned in 2017.

The Compact Accelerator System for Performing Astrophysical Research 
(CASPAR)~\cite{CASPAR-Robertson2016} collaboration is using a 1-MV Van de 
Graaff accelerator to study reactions at stellar energies associated with 
the slow neutron-capture nucleosynthesis process (s-process).  The 
accelerator was relocated from the University of Notre Dame in the summer 
of 2015 to an underground laboratory at the 4850L Ross Campus.  The 
beamline has been assembled and commissioning is underway following first 
beam in May 2017 and an initial operations announcement in July 2017.  
The group expects to begin taking physics data within the next six months.

Black Hills State University has developed an underground campus at the 
4850L Ross Campus.  The main feature of the laboratory is a 74-m$^{2}$ 
cleanroom that hosts multidisciplinary research 
activities~\cite{BHUC-Mount2017}. Space in the cleanest section of the 
laboratory area dedicated to performing low-background assays currently 
hosts five instruments (plus another one outside the cleanroom), for which 
capabilities are summarized in Table~\ref{tab:lbc_sensitivities}.  The 
cleanroom has space to house up to 10 instruments.

\begin{table}[htbp]
\caption{\label{tab:lbc_sensitivities} Low-background counter 
sensitivities for a sample of order $\sim$1 kg and counting for 
approximately two weeks.}
\begin{center}
\begin{tabular}{lcccll}
\br
{\bf Detector} & {\bf Ge}      & {\bf [U]} & {\bf [Th]} & {\bf Install} & {\bf Status} \\
{\bf (Group)}  & {\bf Crystal} & {\bf mBq/kg} & {\bf mBq/kg}  & {\bf Date}    & \\
\mr
Maeve   & 2.2 kg  & 0.1   & 0.1    & BHUC: Nov 2015  & Production \\
(LBNL)  & ($\epsilon$=85\%)  & ($\sim$10~ppt) & ($\sim$25~ppt) & SURF: May 2014 & assays \\[0.2cm]

Morgan  & 2.1 kg  & 0.2   & 0.2    & BHUC: Nov 2015  & Production \\
(LBNL)  & ($\epsilon$=85\%)  & ($\sim$20~ppt) & ($\sim$50~ppt) & SURF: May 
2015 & assays \\[0.2cm]

Mordred      & 1.3 kg & 0.7 & 0.7  & BHUC: Jul 2016 & Production \\
(USD/CUBED,  & ($\epsilon$=60\%) & ($\sim$60~ppt) & ($\sim$175~ppt) & 
SURF: Apr 2013 & assays \\
LBNL)        &&&&& \\[0.2cm]

SOLO       & 0.6 kg & 0.6 & 0.3 & BHUC: Feb 2016 & Production \\
(LZ/UCSB,  & ($\epsilon$=35\%) & ($\sim$50~ppt) & ($\sim$75~ppt) & (from 
Soudan) & assays \\
Brown)     &&&&& \\[0.2cm]

Dual HPGe    & 2$\times$2.1 kg & $\sim$0.01 & $\sim$0.01 & BHUC: Jul 2017 & Commissioning \\
(LBNL,BHSU,  & & ($\sim$1~ppt) & ($\sim$1~ppt) & & \\
UCSB)        &&&&& \\[0.2cm]

Ge-IV      & 2.0 kg  & $<$ 7.4 & $<$ 2.4 & BHUC: Oct 2017 & Initial install \\
(Alabama,  & ($\epsilon$=111\%) & ($<$600~ppt) & ($<$600~ppt) & & \\
USD)       &&&&& \\
\br
\end{tabular}
\end{center}
\end{table}

\section{Future Science}

The future offers many scientific opportunities for underground science at 
SURF.  Upgrades are underway, expansion plans are being developed and 
there exist options to significantly expand the facility footprint to 
accommodate additional endeavors.

The LZ detector~\cite{LZ-TDR2017,LZ-TAUP2017} will employ approximately 
10~tonnes of liquid xenon ($\sim$50$\times$ LUX fiducial) with a projected 
sensitivity 100$\times$ better than the final LUX result. The entire xenon 
inventory is on contract, and the majority has been delivered to SLAC for 
purification. Necessary SURF surface infrastructure upgrades have been 
completed and underground modifications are scheduled to begin later in 
2017, with completion projected by mid-2018. Detector assembly at the 
Surface Laboratory is expected to begin in 2017, with underground 
installation anticipated in 2019 followed by operation in 2020.  The 
nominal data run is 5 years.

The Long Baseline Neutrino Facility (LBNF)/Deep Underground Neutrino 
Experiment (DUNE)~\cite{LBNF_DUNE2015,DUNE-Kemp2017} is the first 
internationally conceived, constructed, and operated mega-science project 
hosted by the Department of Energy in the United States.  Led by Fermilab, 
LBNF will provide facilities at two locations: accelerator facilities at 
Fermilab to create the neutrino beam as well as facilities at SURF to 
support the DUNE detectors that will investigate neutrino properties 
(oscillations, CP violation, mass hierarchy), nucleon decay and supernovae 
using a total of 70~ktonnes (40~kT fiducial) liquid argon on the 4850L. 
Geotechnical studies on the 4850L were completed in the spring of 2014, 
and an initial test-blast program was completed in the spring of 2016.  A 
groundbreaking ceremony for LBNF was held in July 2017, and underground 
construction will begin mid-2018 with the main excavation to commence in 
2019 and last roughly three years.  The current design for the underground 
laboratory envisions four detector chambers, each 20~m wide $\times$ 29~m 
tall $\times$ 70~m long and able to accommodate a 10-kT liquid argon 
detector.  

\section{Summary}

SURF is a deep underground research facility, dedicated to scientific 
uses.  Research activities are supported at a number of facilities, both 
on the surface and underground.  Two campuses on the 4850L accommodate a 
number of leading efforts, and in particular the 4850L Davis Campus has 
been successfully operating for over 5~years.  Several experiments are 
well established and there are robust capabilities for low-background 
counting.  Many expansion possibilities are on the horizon and a number of 
key experiments in the U.S.\ research program are developing plans for 
installation at SURF.


\section*{References}
\bibliography{Heise}

\end{document}